\documentclass[11pt]{article}
\usepackage{moriond,epsfig}

\def\Journal#1#2#3#4{{#1} {\bf #2}, #3 (#4)}

\def\PRL{\em Phys. Rev. Lett.}
\def\PRD{{\em Phys. Rev.} D}

\def\CQG{{\em Class. Quant. Grav.}}
\def\NATURE{{\em Nature}}
\def\AMJP{{\em Am. J. Phys.}}

\def\be{\begin{equation}}
\def\ee{\end{equation}}
\def\bea{\begin{eqnarray}}
\def\eea{\end{eqnarray}}

\begin{document}
\vspace*{4cm}
\title{Significance of the Compton frequency in atom interferometry}

\author{ Michael A. Hohensee and Holger M\"uller }

\address{Department of Physics, University of California, Berkeley, CA 94706, USA}

\maketitle\abstracts{The recent realization that atom interferometers (AIs) can be used to test the gravitational redshift tests has proven to be controversial in some quarters. Here, we address the issues raised against the interpretation of AIs as redshift tests,  reaffirming the fact that M\"uller {\em et al.} [Nature {\bf 463,} 926 (2010)] indeed report a gravitational redshift test.}

\section{Overview}
A variety of arguments have been raised~[2-4] against the interpretation of atom interferometers (AIs) as clock comparisons which test the gravitational redshift~[1].  To this end, non-relativistic derivations of the AI phase have been offered as evidence that the Compton frequency, $mc^{2}/\hbar$, is both unphysical and irrelevant to AI tests of the Einstein Equivalence Principle (EEP)~[2,3].  As we demonstrate below, non-relativistic treatments of the problem obscure the fact that in the fully relativistic theory, matter-waves do indeed oscillate at the Compton frequency, with relevance to tests of the gravitational redshift.  

Other objections focus upon the failings of possible alternative theories of gravity in which critics do consider AIs to be redshift tests~[2,3].  AIs are, it is claimed, more properly understood as tests of the weak equivalence principle (WEP) in any consistent theoretical framework~[2,3].  We address these objections in more detail elsewhere~[5].  For these proceedings, however, we will simply note that in all theories consistent with an action principle and energy conservation, any experiment which constrains violations of WEP also constrains anomalies in the gravitational redshift, and vice versa, as originally hypothesized by Schiff~[6], and covered in detail by~[7-9], as well as more recent reviews of the subject~[2,3].  
The inconsistencies arising from the analysis of AIs in the context of a theory that decouples the gravitational redshift from WEP~[2] result from that very decoupling, and are not specific to any particular experiment.

Critics have also asserted~[2,3] that the precision attained by torsion-balance tests of WEP is such that AI-based redshift tests can only be effectively used to constrain theories that do not conserve energy.  This is ironic, given that in 1975 Nordtvedt~[7] made the same observation regarding the general utility of any clock experiment in the face of the torsion-balance constraints of the day.  The assertion is also incorrect, in light of the subsequent 1977 study by Ni~[10].\footnote{In Ni's theory~[10], the EEP can be violated without generating signals in experiments in which the test masses are restrained from moving freely, as happens in torsion balance ({\em i.e.} ``WEP 1'') experiments.  So far as is currently known, however, the EEP cannot be violated without generating signals in experiments involving freely falling masses (``WEP 2'' tests), AIs, or other clock tests.}

As noted above, non-relativistic treatments of matter-waves obscure the features of the experiment in the relativistic theory. Specifically, the non-relativistic Hamiltonian describing the semiclassical evolution of a particle of mass $m$ in a gravitational potential $U$ takes the form
\begin{equation}
H=mc^{2}+mU(\vec{x})+\frac{{p}^{2}}{2m},\label{eq:nonrelhamone}
\end{equation}
where $\vec{x}$ is the particle's position, and $\vec{p}$ is its momentum.  As the Compton frequency term is constant, the dynamics and accumulated phase of the particle are equivalently modeled by
\begin{equation}
H=mU(\vec{x})+\frac{{p}^{2}}{2m},\label{eq:nonrelhamtwo}
\end{equation}
which some have argued demonstrates that the Compton frequency plays no role~[2,3].  This analysis hides the fact that in the fully relativistic theory, matter-waves do indeed oscillate at the Compton frequency~[11], with important consequences for tests of the EEP.  In what follows, we present a discussion of the significance of the Compton frequency in matter-wave tests of the gravitational redshift, prefaced with a general review of how clock frequencies appear in the derivation of experimental observables in any clock comparison test.  We find that the Compton frequency of matter-waves appears in the same way as does the oscillation frequency of more conventional clocks, and demonstrate that the direct measurement of the clock frequencies is not required to test the gravitational redshift, just as it was unnecessary for the original Pound-Rebka~[12] test.  We close with a gedankenexperiment which further illustrates this equivalence.

\section{Significance of the Clock Frequency to Gravitational Redshift Rests}

All clock comparison tests of the gravitational redshift measure the difference in the phase accumulated by two or more clocks that follow different paths through spacetime.  In the simplest case, all paths originate at a point $A$ and end at a point $B$.  The total phase accumulated by clock $i$ between points $A$ and $B$ is given by
\begin{equation}
\varphi_{(i)}=\omega_{(i)}\int_{A}^{B}d\tau_{(i)}
=\omega_{(i)}\int_{A}^{B}\sqrt{-g_{\mu\nu}dx_{(i)}^{\mu}dx_{(i)}^{\nu}},
\end{equation}
where $\omega_{(i)}$ is the proper oscillation frequency of the $i$th clock, and $x_{(i)}^{\mu}$ denotes the path it takes from $A$ to $B$.  To leading order, this reduces to
\begin{equation}
\varphi_{(i)}=\omega_{(i)}\int_{A}^{B}\left(1+\frac{U(\vec{x}_{(i)}(t))}{c^{2}}
-\frac{\dot{{x}}_{(i)}^{\;2}}{2c^{2}}\right)dt.\label{eq:leadingred}
\end{equation}
The calculated value of $\varphi_{(i)}$ its not itself a physical observable under any circumstances, for any kind of clock.  All physical phase measurements yield the relative phase between two systems.  Thus the physical observable in any clock comparison test is the difference $\Delta\varphi_{ij}=\varphi_{(i)}-\varphi_{(j)}$ between two clocks.  If the clocks have the same proper frequency $\omega_{(i)}=\omega_{(j)}=\omega_{0}$, this becomes
\begin{equation}
\Delta\varphi_{ij}=\omega_{0}\int_{A}^{B}\left(\frac{U(\vec{x}_{(i)}(t))-U(\vec{x}_{(j)}(t))}{c^{2}}-\frac{\dot{{x}}_{(i)}^{\;2}-\dot{{x}}_{(j)}^{\;2}}{2c^{2}}\right)dt.\label{eq:relphisamefreq}
\end{equation}
The first term in the integrand has the form $\Delta U/c^{2}$, and represents the gravitational redshift, while the second term accounts for the phase shift due to time dilation.

Note that we can also obtain Eq.~(\ref{eq:relphisamefreq}) by replacing $\varphi_{(i)}$ and $\varphi_{(j)}$ with the relative phases $\Delta\varphi_{i\infty}$ and $\Delta\varphi_{j\infty}$ that each clock accumulates relative to a fictional oscillator $(\infty)$ lying at rest, far away from sources of the gravitational potential.  Our freedom to designate any oscillator, real or notional, as a reference for the purposes of calculation corresponds to our freedom to add total derivatives to our particle Lagrangian, or equivalently, to choose the zero of our energy scale when transforming Eq.~(\ref{eq:nonrelhamone}) to Eq.~(\ref{eq:nonrelhamtwo}) without changing the physics.  Such transformations can save time in carrying out derivations, but serve to obscure the fundamental oscillation frequency of the clock.  The quantities $\Delta\varphi_{i\infty}$ are no more physically observable than are the $\varphi_{(i)}$, and such a substitution has no impact on the observable phase $\Delta\varphi_{ij}$ accumulated by clocks of any kind.

\section{Significance of the Compton Frequency to Gravitational Redshift Tests\label{sec:compton}}

If we compare Eq.~(\ref{eq:relphisamefreq}) with the phase accumulated by matter-waves propagating along the same paths in the semiclassical limit, we obtain the same result; the phase accumulated by the $i$th matter-wave is simply the integrated action $S_{(i)}/\hbar$.  From the form of the standard general relativistic action for a particle of mass $m$, the phase is given by~[13]
\begin{equation}
\varphi_{(i)}=\frac{S_{(i)}}{\hbar}=\frac{mc^{2}}{\hbar} \int_{A}^{B}d\tau_{(i)},\label{eq:matterphi}
\end{equation}
where $mc^{2}/\hbar$ is the matter-waves' Compton frequency, $\omega_{C}$.
See also~[2,14] for a derivation of Eq.~(\ref{eq:matterphi}) using the Feynman path integral formalism, and of the equivalent Schr\"odinger representation.  As has been pointed out [2,3], the phase $\varphi_{(i)}$ is no more physically observable for matter waves than for any other clock.  The relative phase $\Delta\varphi_{ij}$, however, is observable.  At present, the only practical way to measure the relative phase of two matter-wave oscillators is by interfering coherent superpositions of matter-waves with one another.  Thus $\omega_{(i)}=\omega_{(j)}=\omega_{C}$, and we find that the relative phase is given by Eq.~(\ref{eq:relphisamefreq}) with $\omega_{0}$ replaced by the Compton frequency.

Although the leading order phase shift $\omega_{C}\int dt$ is inaccessible to us,\footnote{In the absence, that is, of a way to coherently convert atoms into microwaves and back again~[2], which would allow us to compare the accumulated matter-wave phase directly to that of a conventional clock.  Note that  $\omega_{0}t$ is never observable in direct measurements of the relative phase between separated yet identical clocks.} this is of no importance for tests of the gravitational redshift, since the Compton frequency also multiplies the redshift and time dilation terms.  It is for this reason that AI tests of the gravitational redshift are so competitive with tests involving conventional clocks, despite the fact that they measure the relative phase with far less precision, operate for far shorter periods of time, and involve clocks separated by much smaller potential differences.

We have shown elsewhere that the specific AI configuration reported in [1] is entirely equivalent to an experiment in which conventional clocks follow the same paths, exchanging signals with a stationary reference clock at discrete intervals~[15].  That the matter-waves' Compton frequency does not necessarily play a role in the derivation of the redshift signal~[2,3] has no bearing on whether the Mach-Zehnder AI constitutes a redshift test.  If it did, one could as easily argue that the Pound-Rebka experiment~[11] is not a redshift test.  There, the redshift of a 14.4 keV M\"ossbauer transition was determined from the velocity $v=\Delta U/c$ at which an identical oscillator must move to compensate for it via the first order Doppler effect.  The actual transition frequency of the 14.4 keV transition drops out of the expression for the velocity, and is not measurable by nor necessary to carry out a test of the redshift.

\section{A Concrete Example\label{sec:gedanken}}
\begin{figure}
\centering
\epsfig{file=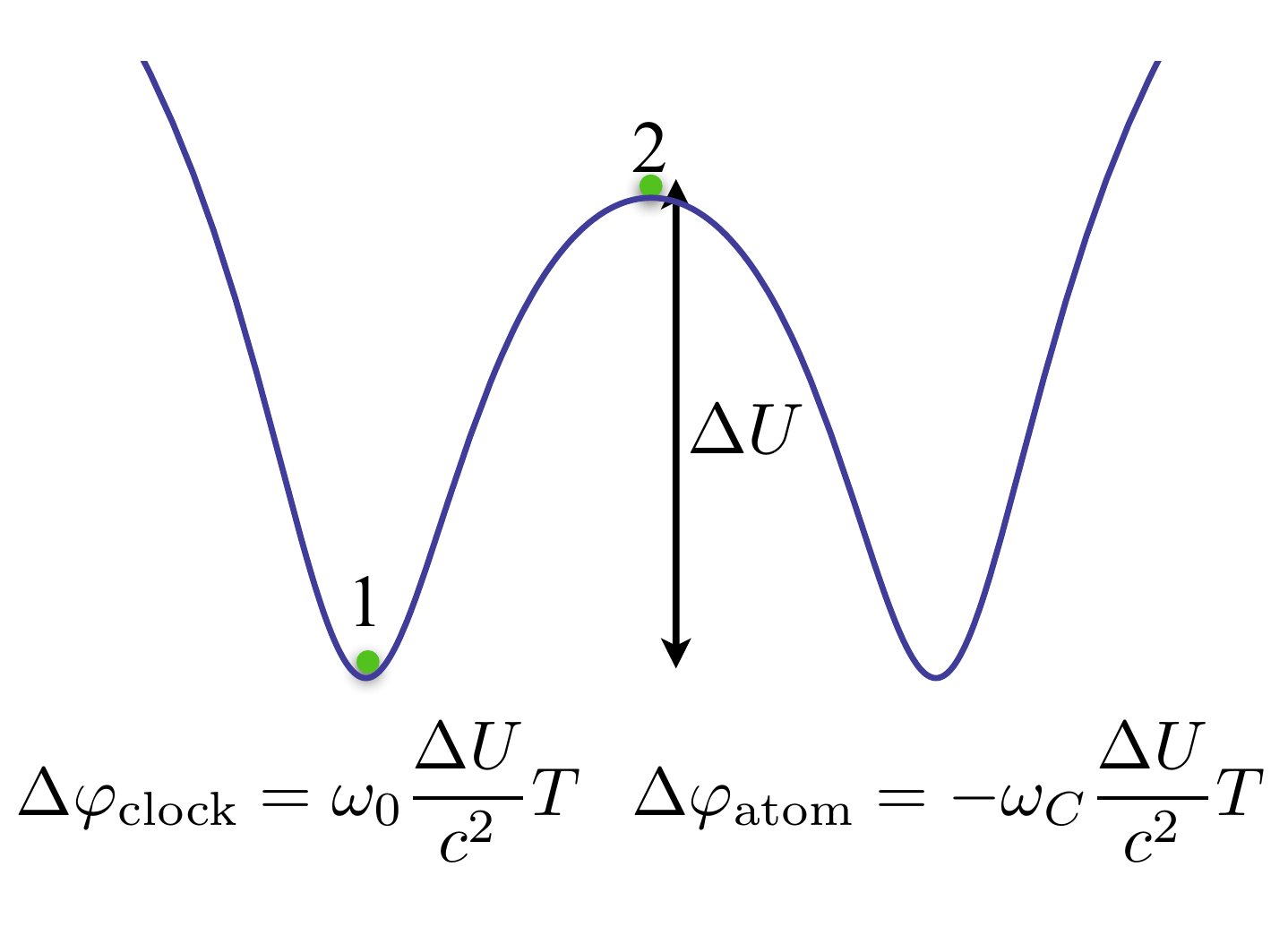,width=0.42\textwidth}
\caption{Gravitational redshift experiment at points of vanishing gravitational acceleration.}
\end{figure}The general situation may be clarified by considering a gedankenexperiment [16], Fig. 1: two halves of a matter-wave are held at the extrema of a gravitational potential, where the local acceleration of free fall is zero. Though the net force acting upon the matter-waves is zero, implying a vanishing gravimeter ({\em i.e.} WEP) signal, they would still accumulate a relative phase $\varphi$ at a rate of $d\varphi /dt=\omega_{C}\Delta U/c^{2}$, as would any pair of similarly positioned clocks ticking with a proper frequency $\omega_{C}$. This follows from derivations in part~\ref{sec:compton}, or from the Schr\"odinger equation.

\section{Conclusion}

Atom interferometers are in every important respect equivalent to other clock tests of general relativity, and their ability to provide competitive limits on violations of EEP stems directly from the fact that the intrinsic oscillation frequency of a matter-wave is the Compton frequency, $mc^{2}/\hbar$.  Matter-wave interferometry enables tests of gravity to be carried out on the tabletop with precision rivaling and in many cases exceeding that of large scale tests with conventional clocks.  AIs close the same loopholes for EEP violation that are addressed by other clock comparison experiments, and have the same experimental characteristics as other clock comparison tests.  AIs will continue to play an essential role in verifying what may be the most important foundational principle of modern physics, and in future searches for physics beyond general relativity and the standard model.

\section*{Acknowledgments}
We are grateful to S. Chu, P. Hamilton, and A. Zeilinger for discussions, and the David and Lucile Packard Foundation, the National Institute of Standards and Technology, the National Aeronautics and Space Administration, and the Alfred P. Sloan Foundation for support.

\end{document}